\documentclass[sn-nature]{sn-jnl}

\usepackage{graphicx}%
\usepackage{multirow}%
\usepackage{amsmath,amssymb,amsfonts}%
\usepackage{amsthm}%
\usepackage{mathrsfs}%
\usepackage[title]{appendix}%
\usepackage{xcolor}%
\usepackage{textcomp}%
\usepackage{manyfoot}%
\usepackage{booktabs}%
\usepackage{tabularx} 
\usepackage{algorithm}%
\usepackage{algorithmicx}%
\usepackage{algpseudocode}%
\usepackage{array}
\usepackage{listings}%
\usepackage{tikz}
\usetikzlibrary{positioning, shapes.multipart, arrows.meta}
\usepackage[edges]{forest}
\usepackage{float}
\usepackage{hyperref}
\usepackage{arydshln}
\usepackage{multirow}

\theoremstyle{thmstyleone}%
\theoremstyle{thmstyletwo}%

\theoremstyle{thmstylethree}%

\usepackage{geometry}
\begin{document}

\title[Article Title]{Meta-Analysis of Transfer Learning for Segmentation of Brain Lesions}


\author[1,2]{\fnm{Sovesh} \sur{Mohapatra}}

\author[2]{\fnm{Advait} \sur{Gosai}}

\author[3,5]{\fnm {Anant} \sur{Shinde}}

\author[4] {\fnm{Aleksei} \sur{Rutkovskii}}

\author[5]{\fnm {Sirisha} \sur{Nouduri}}

\author*[1,3,5]{\fnm{Gottfried} \sur{Schlaug}}\email{gschlaug@umass.edu}

\affil[1]{\orgdiv{Institute for Applied Life Sciences}, \orgname{University of Massachusetts}, \orgaddress{\city{Amherst}, \postcode{01003}, \state{MA}}}

\affil[2]{\orgdiv{Manning College of Information and Computer Sciences}, \orgname{University of Massachusetts}, \orgaddress{\city{Amherst}, \postcode{01003}, \state{MA}}}

\affil[3]{\orgdiv{Department of Biomedical Engineering}, \orgname{University of Massachusetts}, \orgaddress{\city{Amherst}, \postcode{01003}, \state{MA}}}

\affil[4]{\orgdiv{Department of Electrical and Computer Engineering}, \orgname{University of Massachusetts}, \orgaddress{\city{Amherst}, \postcode{01003}, \state{MA}}}

\affil[5]{\orgdiv{Department of Neurology}, \orgname{Baystate Medical Center, and UMass Chan Medical School - Baystate Campus}, \orgaddress{\city{Springfield}, \postcode{01107}, \state{MA}}}


\abstract{A major challenge in stroke research and stroke recovery predictions is the determination of a stroke lesion’s extent and its impact on relevant brain systems. Manual segmentation of stroke lesions from 3D magnetic resonance (MR) imaging volumes, the current gold standard, is not only very time-consuming, but its accuracy highly depends on the operator's experience. As a result, there is a need for a fully automated segmentation method that can efficiently and objectively measure lesion extent and the impact of each lesion to predict impairment and recovery potential which might be beneficial for clinical, translational, and research settings. We have implemented and tested a fully automatic method for stroke lesion segmentation which was developed using eight different 2D-model architectures  trained via transfer learning (TL) and mixed data approaches. Additionally, the final prediction was made using a novel ensemble method involving stacking and agreement window. Our novel method was evaluated in a novel in-house dataset containing 22 T1w brain MR images, which were challenging in various perspectives, but mostly because they included T1w MR images from the subacute (which typically less well defined T1 lesions) and chronic stroke phase (which typically means well defined T1-lesions). Cross-validation results indicate that our new method can efficiently and automatically segment lesions fast and  with high accuracy compared to ground truth. In addition to segmentation, we provide lesion volume and weighted lesion load of relevant brain systems based on the lesions' overlap with a canonical structural motor system that stretches from the cortical motor region to the lowest end of the brain stem. Such a combination of an automatically determined lesion with its impact on a relevant brain system provides a fast and objective, user-experience independent functional understanding of a lesions’ extent and impact.}




\maketitle 

\section{Introduction}
\label{sec:introduction}

Strokes are a leading cause of long-term disability and mortality rates, impacting millions of individuals annually in the world \cite{feigin2021global}. In the United States alone, around 795,000 people experience either a new or a recurrent stroke each year, with approximately 610,000 being first-time occurrences and 185,000 being recurrent attacks \cite{tsao2022heart, tsao2023heart}. The identification and segmentation of a stroke lesion plays an important role for accurate diagnosis and etiology, prognosis, and treatment options \cite{atkins1998fully}. With extensive and long-time training, clinical and translational professionals can define the extent of a lesion through visual inspection, however, accurate manual segmentation and determining the impact of a lesion on relevant and eloquent regions of the brain is time-consuming, expensive, and can be error-prone due to the expertise necessary for the visual analysis of images and to keep the inter-observer variability to a minimum \cite{wilke2011manual}. Thus, developing user independent methods for rapid determination of any lesion, in particular lesions in both subacute and chronic phase of a stroke, becomes an important endeavor in clinical and research settings, allows the stratification of patients into clinical trials, and enhances our understanding of a lesions' impact on a patient’s functional impairment and their stroke outcome potential \cite{zhang2018automatic, hu2020brain, yu2023san}.

Transfer learning (TL) uses pre-trained models to extract significant information and features from one domain and apply them to a related but distinct separate domain. In the context of medical imaging, TL has emerged as an effective approach to deal with issues including the lack of labeled data and the need for robust feature extraction \cite{raghu2019transfusion}. Researchers can initiate their work with models that have already been pre-trained on extensive, general-purpose datasets, possessing a feature set capable of identifying and capturing significant attributes. The models can then be fine-tuned for the particular medical imaging needs at hand or the particular imaging dataset requiring segmentation \cite{tajbakhsh2016convolutional, shin2016deep}. Intermediate task training (ImTT), a subset of TL, involves training a model on a series of related tasks before fine-tuning on the target task \cite{he2021hf, kang2022thyroid}. Recent studies have demonstrated the efficacy of TL and ImTT in various medical imaging tasks, including lesion segmentation \cite{gryska2021automatic}, tumor classification \cite{shin2016deep}, and tissue compartments  identification \cite{zhu2019anatomynet}.

To explore the potential of TL in 2D based deep learning models, we conducted a comprehensive meta-analysis focusing on the index lesion segmentation in stroke patients that had imaging studies obtained in the subacute and chronic phase after an index stroke event had occurred. The emphasis on the index lesion in our paper was done with the understanding that there are other abnormal regions in a brain with a stroke that might be due to chronic small vessel disease related to the presence of long-standing vascular risk factors (e.g., smoking, high blood pressure, hyperlipidemia, diabetes, etc)  or due to secondary degeneration of white matter tracts induced by the index lesion and more visible in the subacute and chronic stroke phase. Our study compared the performance of models utilizing ImTT with those using mixed data. Furthermore, we explored the effectiveness of ensembling fine-tuned models trained on mixed data by incorporating a binary overlap (stacking) and a novel window approach. Our study also emphasized the fact that through these techniques, 2D models can achieve comparable accuracy to hand-drawn lesion masks while requiring fewer computational resources.

\section{Methodology} \label{methods}

\subsection{Selection of MRI data} \label{select mri}

We examined 602 anonymized MR brain images and their corresponding hand-drawn lesions maps that resulted from a consensus of multiple experienced raters curated from Anatomical Tracings of Lesions After Stroke (ATLAS) dataset \cite{liew2018large} (547 high resolution T1w MR brain images) and our anonymized and de-identified in-house dataset (55 high resolution T1w MR brain images, either 1X1X1 acquisition voxel size or resampled to 1X1X1 voxel size). This encompassed a wide range of acute, subacute and chronic stroke patients (as defined by \cite{liew2018large}) lesions in various brain regions as defined in Section \ref{split}

\subsection{Preprocessing of MRI data} \label{preprocessing}

Several preprocessing measures were conducted to prepare the MRI data for further analysis. These steps were necessary to standardize the data and enhance the performance of our deep learning models. Firstly, all MRI images were resampled to a consistent size of 256x256x128 voxels to ensure uniformity across the dataset. The skull, meninges, and ventricles were then removed/excluded from the images. Subsequently, the images and their corresponding masks were sliced in the axial plane to generate appropriate input to our 2D models. 
To enhance the robustness of our model, various image augmentation techniques were applied to generate multiple variants of the original images. This approach increases the dataset size and introduces variability, which ultimately improves the model's ability to generalize and perform well on unseen data. The augmentation process was carried out using a custom Python function leveraging the ImageDataGenerator class from the Keras library \cite{chollet2016building}. Finally, all images and masks were normalized with the mask values being binarized.

\subsection {Splitting the whole brain into multiple regions} \label {split}

Based on our comprehensive understanding of anatomy, major vascular systems, and clinical knowledge regarding a stroke lesion’s impact of particular brain systems (the motor system in our dataset), we divided the whole brain into four different \emph{“super-regions”}, which were gross-anatomically defined. Dividing the brain into these major \emph{super-regions} allowed us to identify which model architectures learned and captured different unique complexities associated with each region.  \\ 
Super Region 1 (R1) consists of the inferior frontal region and sub-cortical deep gray matter regions of the brain (e.g., basal ganglia, thalamus, insular cortex) \\
Super Region 2 (R2) consists of the cerebellum and three parts of the brainstem (medulla, pons, and midbrain regions). \\
Super Region 3 (R3) consists of the superior part of the frontal lobe and the entire occipital lobe and parietal lobe.\\ 
Super Region 4 (R4) consists of the temporal lobe, limbic lobe (e.g., cingulate gyrus), and fronto-mesial part of the brain.

\begin{figure}[h]
    \centering
    \includegraphics[width=\linewidth]{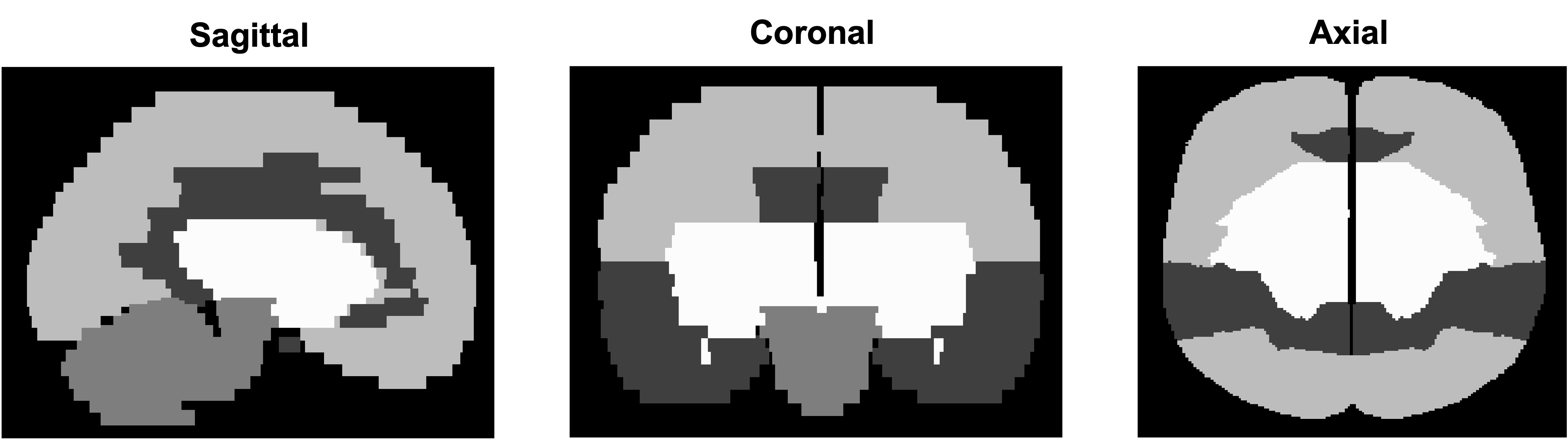}
    \caption{MNI152 1mm brain with separate \emph{super-regions} marked in different signal intensities}
    \label{1_x}
\end{figure}

All the sub-regions in each of the \emph{super-regions} were selected from the Talairach atlas \cite{lancaster1997automated, lancaster2000automated} as shown in Figure \ref{1_x}. The subregions were added and multiplied with consequent intensity values using the FSL software. As shown in Section \ref{ensembleres}, the models' performance was improved by this approach when ensemble methods were applied.

\subsection{Model training}\label{mt}

We conducted an extensive study by training and evaluating eight distinct model architectures to enhance the analysis. These architectures encompass a wide variety of approaches, including U-Net, U-Net++, Residual U-Net (ResUNet), Residual Network (ResNet), VGG U-Net (VGGUNet), V-Net, Fully Convolutional DenseNet (FC DenseNet), and Attention U-Net (Att UNet). All models were trained on a high-performance computer equipped with 128 gigabytes of RAM, an NVIDIA GeForce RTX 3080 GPU with 12 gigabytes of memory, and an AMD Threadripper Pro 5955WX Processor, running on the Ubuntu 20.04 operating system. Detailed information about the hyperparameters and model architectures can be found in the supporting information (SI) section of our manuscript.

Our meta-analysis consisted of primarily comparing two different model training and fine-tuning approaches: training and validation on mixed data and training using TL with ImTT. Our aim was to compare the efficacy of both approaches for the specific task of subacute/chronic stroke lesion segmentation.  

\subsubsection{Mixed data}

We trained, validated and tested the models on slices of all 547 MR images along with their corresponding hand-drawn lesion masks from the ATLAS v2.0 dataset and 33 of the image mask pairs from our in-house dataset. In this case, the training set was made up of 80\% of all 2D slices of all MR image datasets, while the validation set (used for hyperparameter tuning) and testing set each contained 10\%. Finally, the models were evaluated on the independent test set consisting of the remaining 22 MR brain images from the in-house dataset.

\subsubsection{Intermediate Task Training}

We initially trained the models on slices of 447 MR images along with their corresponding hand-drawn lesion masks curated from the ATLAS v2.0 dataset. Next, we divided the remaining 100 MR images from the ATLAS into four groups based on the lesion locations, as depicted in Section \ref{split}. We then fine-tuned the models with the brains and corresponding hand-drawn lesion masks focusing on specific \emph{super-regions}, further fine-tuning them on our target task which is segmenting our in-house dataset by using 33 of its T1w MR images. These models too were evaluated on the independent test set as described previously. The suffix \_FT indicates models trained using ImTT.

\begin{figure}[h]
    \centering
    \includegraphics[width=0.8\linewidth]{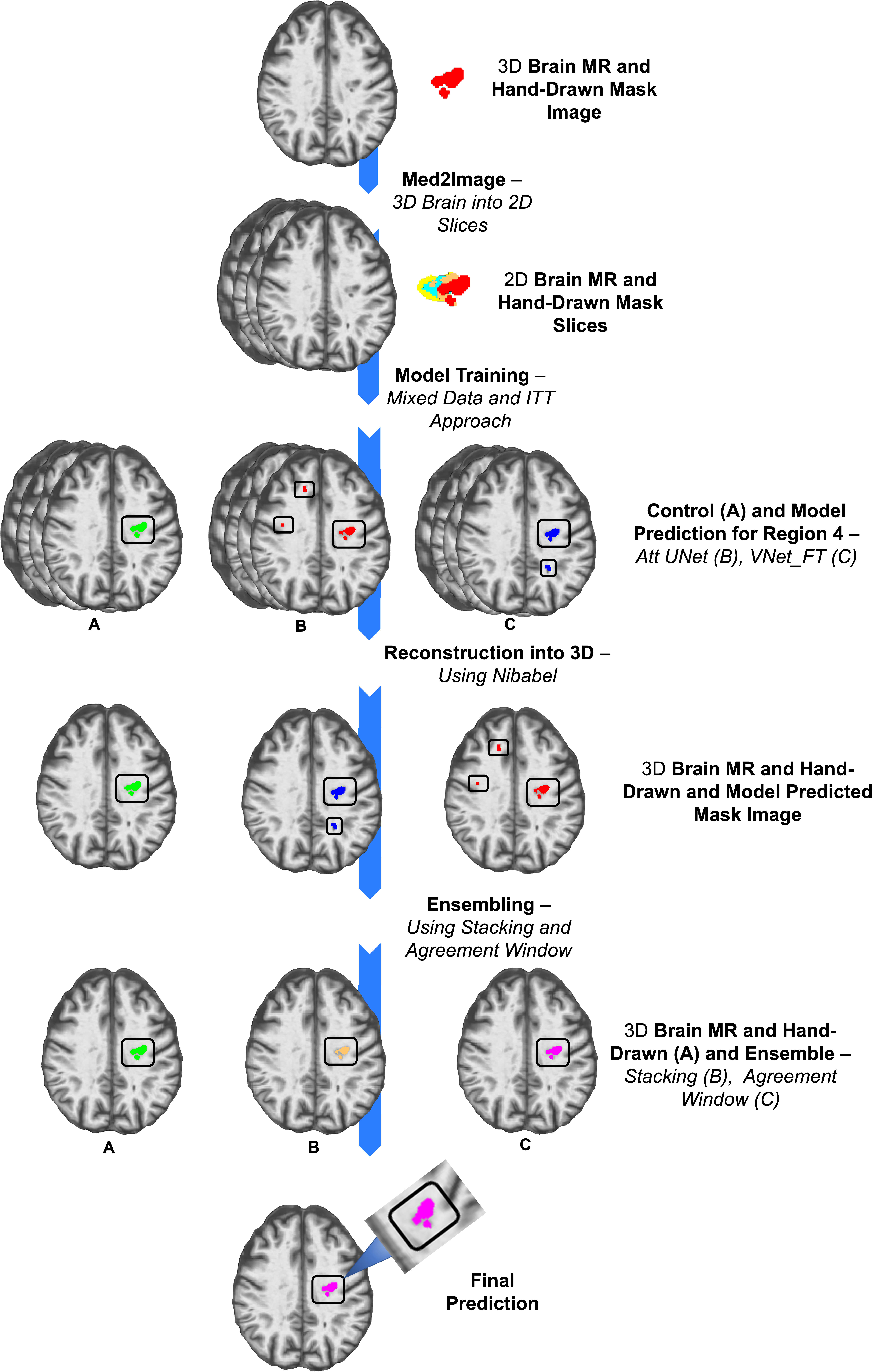}
    \caption{Comprehensive workflow – visualizing the end-to-end process with model training and ensembling approaches}
    \label{fig:1}
\end{figure}

\subsection{Ensembles of trained models}

In addition to using two different training approaches, we examined different ensembles of these eight unique model architectures to evaluate whether a collective strategy would produce superior results compared to relying on a single model after applying the binary mask to the predicted lesion masks. We implemented two ensemble techniques as detailed below.

\subsubsection{Stacking}

The stacking technique employed in this study utilized a binary overlap method. The final mask for each image was determined by considering the overlapping voxels resulting from stacking individual model predictions. To construct the stacks, we chose the four top-performing models per \emph{super-region}, and analyzed all of their possible combinations to derive the best ensemble. The stacking method has been illustrated in Figure \ref{fig:1}.

\subsubsection{Agreement Window}

The agreement window ensemble method is a novel approach developed for this study that combined a window kernel with stacking. The 3D window kernel was designed to retain the union of predicted voxels if their stacking overlap was greater than a certain threshold; whereas it would discard all voxels in the window if the minimum overlap was not satisfied. This window was convolved on the chosen stack of individual model predictions to generate a final outcome. While binary stacking generally resulted in a smaller, less descriptive mask, this method allowed us to eliminate predicted noise from individual models, as well as retain a more accurate mask shape around the targeted index lesion area. Different values of the window size and overlap threshold were tested to identify the optimal parameters. The agreement window method is  illustrated in Figure \ref{fig:1}.

\subsection{Evaluation metrics}

We assessed our results using four distinct metrics: dice coefficient, Jaccard Index (also known as IoU), precision, and recall to gain insight into the performance of the models and ensemble methods. Detailed definitions of each of the chosen evaluation metrics and their relevance can be found in the SI.

\subsection{Evaluating lesion impact on motor regions}

We employed a systematic approach to assess the impact of brain lesions in stroke patients on relevant or eloquent canonical brain structures, such as the corticospinal tract (CST), connecting the motor cortex with alpha motor neurons in the spinal cord. The canonical CST should be understood as a surrogate structural marker of the extent, intersubject variability as well as anatomical variability of a relevant motor system \cite{zhu2010lesion, feng2015corticospinal}. Based on the CST-lesion load value for either side of the brain \cite{zhu2010lesion}, we classified the impacts into three categories on the basis of incremental order of lesion load values such as Small, Medium, and Large.

\begin{itemize}
    \item Large: Indicates a high lesion load, corresponding to a severe neurological impairment.

  \item Medium: Indicates a medium lesion load, corresponding to a moderate neurological impairment.

    \item Small: Indicates a small lesion load, corresponding to a mild neurological impairment .

\end{itemize}

This evaluation parameter not only helps in understanding the anatomical impacts of lesions on a relevant brain system but also examines the reliability of the model predictions as detailed in Section \ref{final}.

\section{Results}

\subsection{Performance assessment of 2D-based DL models}

\begin{table}[h]
    \centering
    \caption{Comparison of performance of models segmenting the lesions in the whole brains of independent validation set. (DSC = Dice Similarity Coefficient, Prec = Precision, Rec = Recall, ER LV = \% Error Rate of Lesion Volume, ER LL = \% Error Rate of Lesion Load)}
    \label{tab:1}
    \addtolength{\tabcolsep}{-3pt}
    \begin{tabular}{ccccccccccccc}
        \hline \\[-1.5ex]
        \multicolumn{1}{c}{} & \multicolumn{6}{c}{\textbf{Mixed Data}} & \multicolumn{6}{c}{\textbf{Intermediate Task Training}} \\[1ex]
        \textbf{Model} & DSC & IoU & Prec & Rec & ER LV & ER LL & DSC & IoU & Prec & Rec & ER LV & ER LL \\[1ex]
        \hline \\[-1.5ex]
        Att UNet & 0.647 & 0.497 & \textbf{0.754} & 0.584 & 64.65 & 23.07 & \textbf{0.701} & \textbf{0.552} & 0.656 & \textbf{0.789} & \textbf{35.74} & \textbf{22.91} \\ [1ex]
        FC DenseNet & 0.580 & 0.442 & 0.600 & 0.602 & 75.30 & 44.64 & \textbf{0.715} & \textbf{0.564} & \textbf{0.711} & \textbf{0.738} & \textbf{41.43} & \textbf{27.70} \\ [1ex]
        ResNet & 0.623 & 0.467 & \textbf{0.772} & 0.545 & 87.87 & 46.53 & \textbf{0.661} & \textbf{0.523} & 0.590 & \textbf{0.807} & \textbf{38.34} & \textbf{30.09} \\ [1ex]
        ResUNet & 0.622 & 0.434 & 0.601 & 0.646 & 99.33 & 46.22 & \textbf{0.654} & \textbf{0.514} & \textbf{0.631} & \textbf{0.748} & \textbf{72.18} & \textbf{33.58} \\[1ex]
        UNet & 0.602 & 0.455 & \textbf{0.609} & 0.679 & 89.00 & 36.29 & \textbf{0.607} & \textbf{0.461} & 0.572 & \textbf{0.708} & \textbf{81.70} & \textbf{35.24} \\[1ex]
        UNet++ & \textbf{0.686} & \textbf{0.597} & 0.653 & \textbf{0.695} & \textbf{30.12} & \textbf{23.28} & 0.633 & 0.512 & \textbf{0.655} & 0.629 & 67.83 & 25.35 \\[1ex]
        VGG UNet & 0.570 & 0.414 & \textbf{0.628} & 0.583 & 74.56 & 41.16  & \textbf{0.615} & \textbf{0.473} & 0.590 & \textbf{0.702} & \textbf{60.22} & \textbf{32.63} \\[1ex]
        VNet & \textbf{0.711} & \textbf{0.569} & 0.697 & \textbf{0.773} & 62.60  & \textbf{24.16} & 0.700 & 0.552 & \textbf{0.755} & 0.685 & 49.00 & 25.28 \\[1ex]
        \hline
    \end{tabular}
\end{table}

Based on our evaluation of the models using an independent set of 22 T1w MR Sequences from our in-house dataset and the chosen evaluation metrics, we observed that the ImTT training approach is more effective than the mixed-data approach, particularly when considering the entire brain. Additionally, the evaluation of predictions from a single model revealed that the Att-UNet\_FT model outperforms all other models in terms of the DSC, IoU and ER LL metrics while the ResNet\_FT model has the lowest error rate in predicting the Lesion Volume as shown in Table \ref{tab:1}. 

\subsection{Performance assessment of model prediction in \emph{super-region-wise} and ensemble methods} \label{ensembleres}

In comparison to performing a whole-brain analysis, we identified four anatomically  distinct brain \emph{super-regions} and evaluated the performance of individual models and ensemble methods in calculating critical lesion information - lesion volume and lesion load. 

Following the labeling of stroke locations in our training data with their corresponding brain \emph{super-regions}, we selected the top 4 performing models for each \emph{super-region} and applied the two ensemble methods as described in Section \ref{methods}. In our analysis, we combined \emph{super-regions} 1 and 4, as we observed that all lesions present in \emph{super-regions} 1 also overlapped with \emph{super-regions} 4. 

Table \ref{tab:2} presents the \emph{super-region-wise} individual model and ensemble performances. It is observed that the agreement window ensemble method consistently produces lower error rates than the stacking method for all \emph{super-regions}, while only being outperformed by the Att UNet\_FT model for strokes present in \emph{super-regions} 3. Specifically, when evaluated based on lesion load, the agreement window predictions have significantly improved and have a lower average error rate across all \emph{super-regions}, as opposed to the average error rates from individual models and binary stacking. 

\begin{table}[h]
\centering
\caption{Comparison of ER LV and ER LL across different models’ \emph{super-region-wise} predictions and ensemble approaches. Top 4 models per \emph{super region(s)} were chosen; all combinations of which were then ensembled, out of which the two best combinations are reported. The optimal (window size, overlap threshold) parameters found empirically for the agreement window approach are denoted next to the respective model names. }
\label{tab:2}
\begin{tabular}{ccccc}
\hline \\[-1.5ex]
\textbf{Super Region(s}) & \textbf{Method} & \textbf{Top Model(s)} & \textbf{ER LV} & \textbf{ER LL} \\ [1.5ex] \hline \\[-1.5ex]
\multirow{8}{*}{1 \& 4} & \multirow{4}{*}{Individual} & Att UNet\_FT & 139.97 & 18.41 \\
 &  & FC DenseNet\_FT & 188.51 & 15.04 \\
 &  & ResNet\_FT & 51.16 & 17.40 \\
 &  & UNet$++$ & 189.52 & 14.90 \\ [0.5ex]
 & \multirow{2}{*}{Stack} & FC DenseNet\_FT, UNet$++$ & 43.37 & 14.76 \\
 &  & FC DenseNet\_FT, ResNet\_FT & 46.52 & 15.92 \\ [0.5ex]
 & \multirow{2}{*}{Agreement Window} & FC DenseNet\_FT, UNet$++$ (2, 0.75) & 38.37 & \textbf{7.73} \\
 &  & FC DenseNet\_FT, UNet$++$ (2, 0.75) & \textbf{33.45} & 8.42 \\
[1.5ex] \hline \\[-1.5ex]
\multirow{8}{*}{2} & \multirow{4}{*}{Individual} & Att UNet & 99.03 & 13.83 \\
 &  & ResNet\_FT & 51.48 & 20.46 \\
 &  & ResUNet & 82.83 & 16.33 \\
 &  & VNet & 50.91 & 15.79 \\ [0.5ex]
 & \multirow{2}{*}{Stack} & Att UNet, ResUNet & 29.95 & 20.83 \\
 &  & Att UNet, VNet & 28.25 & 20.04 \\ [0.5ex]
 & \multirow{2}{*}{Agreement Window} & Att UNet, VNet (3, 0.5) & 25.81 & 12.56 \\
 &  & Att UNet, VNet (4, 0.5) & \textbf{26.62} & \textbf{13.89} \\
[1.5ex] \hline \\[-1.5ex]
\multirow{8}{*}{3} & \multirow{4}{*}{Individual} & Att UNet & 16.15 & 18.14 \\
 &  & Att UNet\_FT & \textbf{11.78} & \textbf{11.09} \\
 &  & UNet++ & 14.52 & 16.21 \\
 &  & VNet & 27.65 & 16.10 \\ [0.5ex]
 & \multirow{2}{*}{Stack} & Att UNet, VNet & 13.92 & 20.64 \\
 &  & UNet++, VNet & 11.86 & 15.12 \\ [0.5ex]
 & \multirow{2}{*}{Agreement Window} & UNet++, VNet (2, 0.75) & 11.72 & 14.65 \\
 &  & UNet++, VNet (3, 0.5) & 10.89 & 15.01 \\ 
[1.5ex] \hline
\end{tabular}
\end{table}

Table \ref{tab:3} presents a comprehensive comparison of the top-performing model(s) for all of the different whole-brain and \emph{super-region-wise} ensemble methods described in this paper. The \emph{super-region-specific} predictions of the top performing models from each \emph{super-region} (as shown in Table \ref{tab:2}) were aggregated to generate a final \emph{super-region-wise} lesion prediction for a target brain.

Examining the results, we notice that \emph{super-region-wise} ensembles surpass the performance of whole-brain predictions by individual models and even whole-brain ensembles. Furthermore, the effectiveness of the agreement window method is seen by its superior performance in both the whole-brain and \emph{super-region-wise} approaches, with the \emph{super-region-wise} agreement window method generating the best lesion volume and lesion load predictions. 

\begin{table}[h]
\centering
\caption{Comparison of top models and their ensembles based on all of the techniques presented in the study. (AW = Agreement Window)}
\label{tab:3}
\addtolength{\tabcolsep}{-2pt}
\begin{tabular}{cccccc}
\hline \\[-1.5ex]
\textbf{Scope} & \textbf{Method}                                            & \textbf{Top Model(s)} & \textbf{DSC} & \textbf{ER LV} & \textbf{ER LL} \\ [1.5ex] \hline \\[-1.5ex]
Whole-Brain    & Mixed Data                                                  & UNet++                                                                                                                                              & 0.686 & 30.12          & 23.28          \\ [1.5ex]
Whole-Brain    & ImTT                                                        & Att UNet\_FT                                                                                                                                    & 0.701    & 35.74          & 22.91          \\ [1.5ex]
Whole-Brain    & Stack                                                       & Att UNet\_FT, UNet++                                                                                                                        & 0.720        & 41.34          & 26.93          \\ [1.5ex]
Whole-Brain   & \begin{tabular}[c]{@{}c@{}}AW \end{tabular} & Att UNet\_FT, UNet++ (3, 0.5)                                                                                                          & 0.726         & 41.25          & 22.11          \\ [1.5ex]
Super-Region-wise    & Stack                                                       & \begin{tabular}[c]{@{}c@{}}{[}R1, R4{]}: FC DenseNet\_FT, UNet++\\ {[}R2{]}: AttUNet, VNet\\ {[}R3{]}: AttUNet\_FT\end{tabular}        & 0.719        & 37.87          & 23.44          \\ [3.5ex]
Super-Region-wise    & \begin{tabular}[c]{@{}c@{}}AW \end{tabular} & \begin{tabular}[c]{@{}c@{}}{[}R1, R4{]}: FC DenseNet\_FT, UNet++ (3, 0.75)\\ {[}R2{]}: AttUNet, VNet (3, 0.5)\\ {[}R3{]}: AttUNet\_FT\end{tabular} & \textbf{0.736} & \textbf{25.58} & \textbf{15.47} \\ [3.5ex] \hline
\end{tabular}
\end{table}

\subsection{Performance assessment of final predictions using lesion load} \label{final}

\begin{figure}[h]
    \centering
    \includegraphics[width=\linewidth]{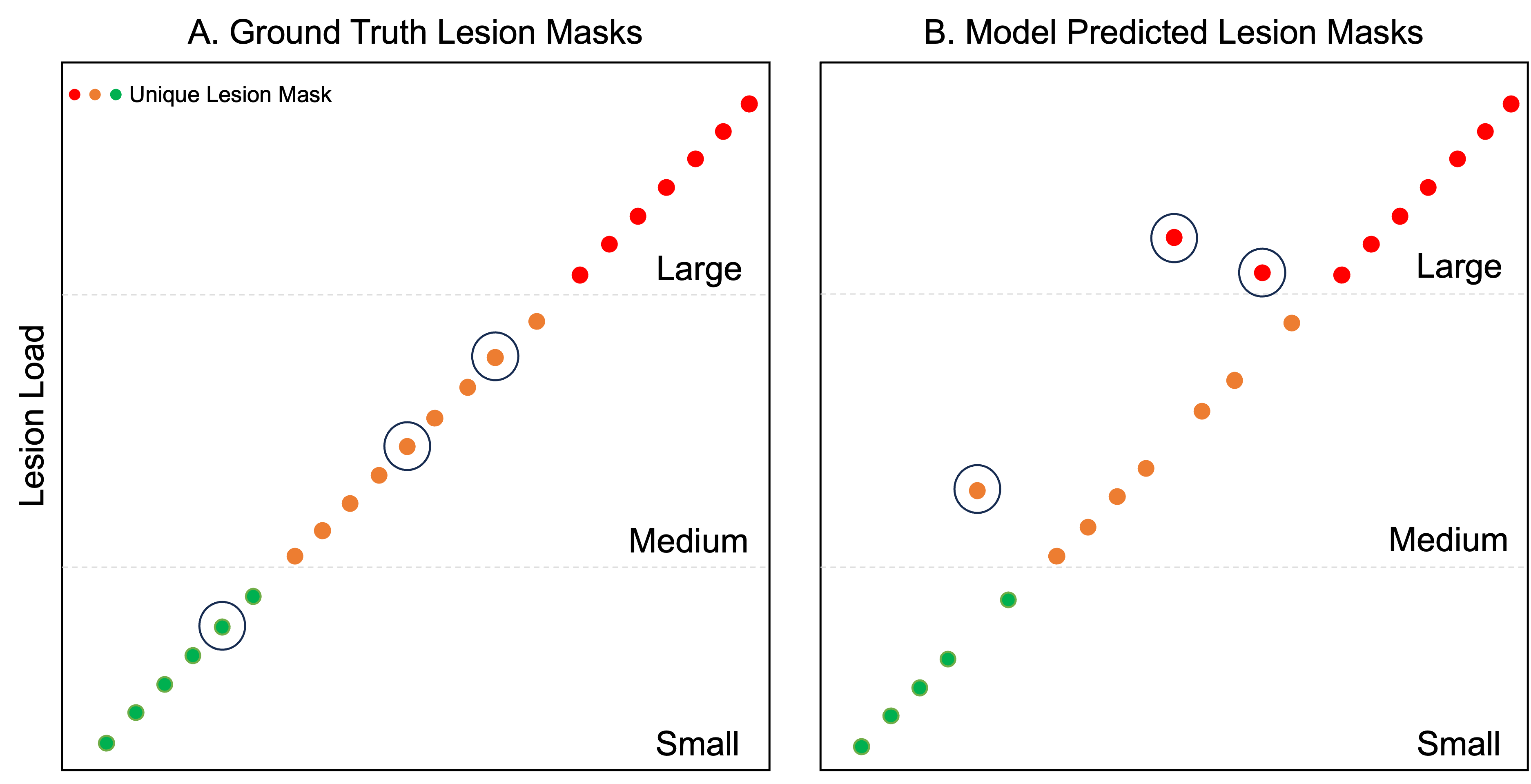}
    \caption{Comparison between hand-drawn and model-predicted lesion masks, including severity levels on motor region based on lesion load value. The x-axis are the individual subjects aligned according to their ground truth and model predicted CST-lesion load (from left to right in ascending order). The circles in \textbf{B} denote outliers or inaccurate predictions made by the model.}
    \label{fig:2}
\end{figure}

Figure \ref{fig:2} illustrates the comparison between the three lesion categories of hand-drawn lesion masks and predicted lesion masks, based on their respective CST lesion load values. Out of 22 predictions, 18 exhibit category  levels within the same range, and most instances have a significantly low error rate. The overall error rate for the models used is presented in Table \ref{tab:2}.

Figure \ref{fig:2} clearly shows three instances of outliers or inaccurate predictions in the final outcomes. Across all three cases, the model excessively estimated the CST lesion load values, incorrectly moving one from the small to medium range and two from the medium to large. This trend signifies an overestimation of lesion by the model, which can be viewed more favorably than if the model has an underestimation of lesion, as the overestimation ensured that the true lesions were not left out.

\begin{figure}[h]
    \centering
    \includegraphics[width=\linewidth]{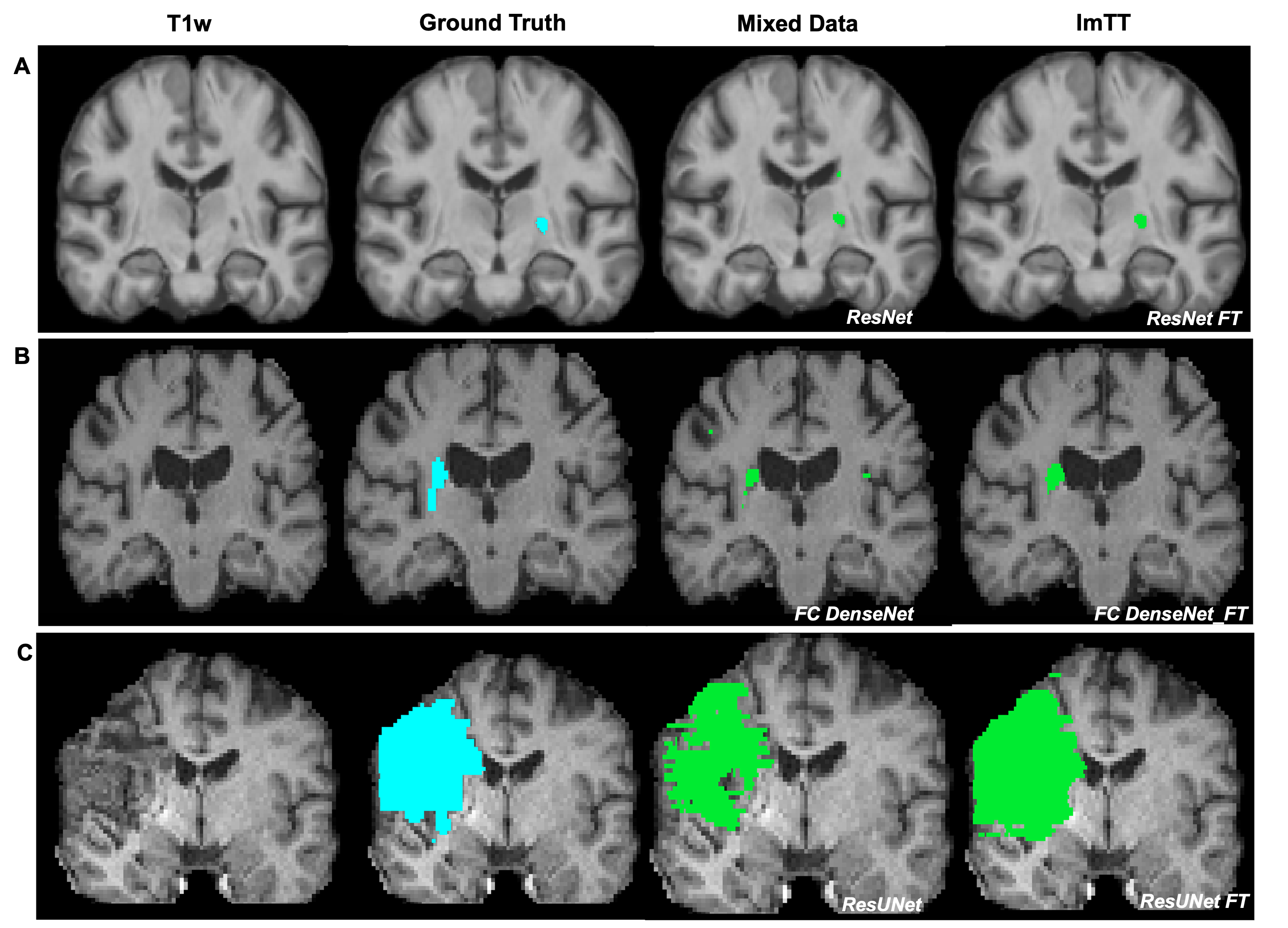}    
    \caption{Comparison of hand-drawn lesion maps and lesion map predictions from two distinct model training approaches. \textbf{A, B, C} display predictions from ResNet, FC DenseNet, and ResUNet (trained with mixed data) alongside ResNet\_FT, FC DenseNet\_FT, and ResUNet\_FT (trained using ImTT) respectively.}
    \label{fig:3}
\end{figure}

\subsection{Visual comparison}

Figure \ref{fig:3} presents a comparative analysis of prediction results from two distinct training approaches for models, as discussed in Section \ref{mt}. \textbf{A, B, C} shows three separate predictions generated by three different model architectures on brain MR images in the independent validation set. It is evident that the mixed data approach leads to some false positives in the models' predictions, which are effectively eliminated when employing the ImTT training approach.

\begin{figure}[h]
    \centering
    \includegraphics[width=\linewidth]{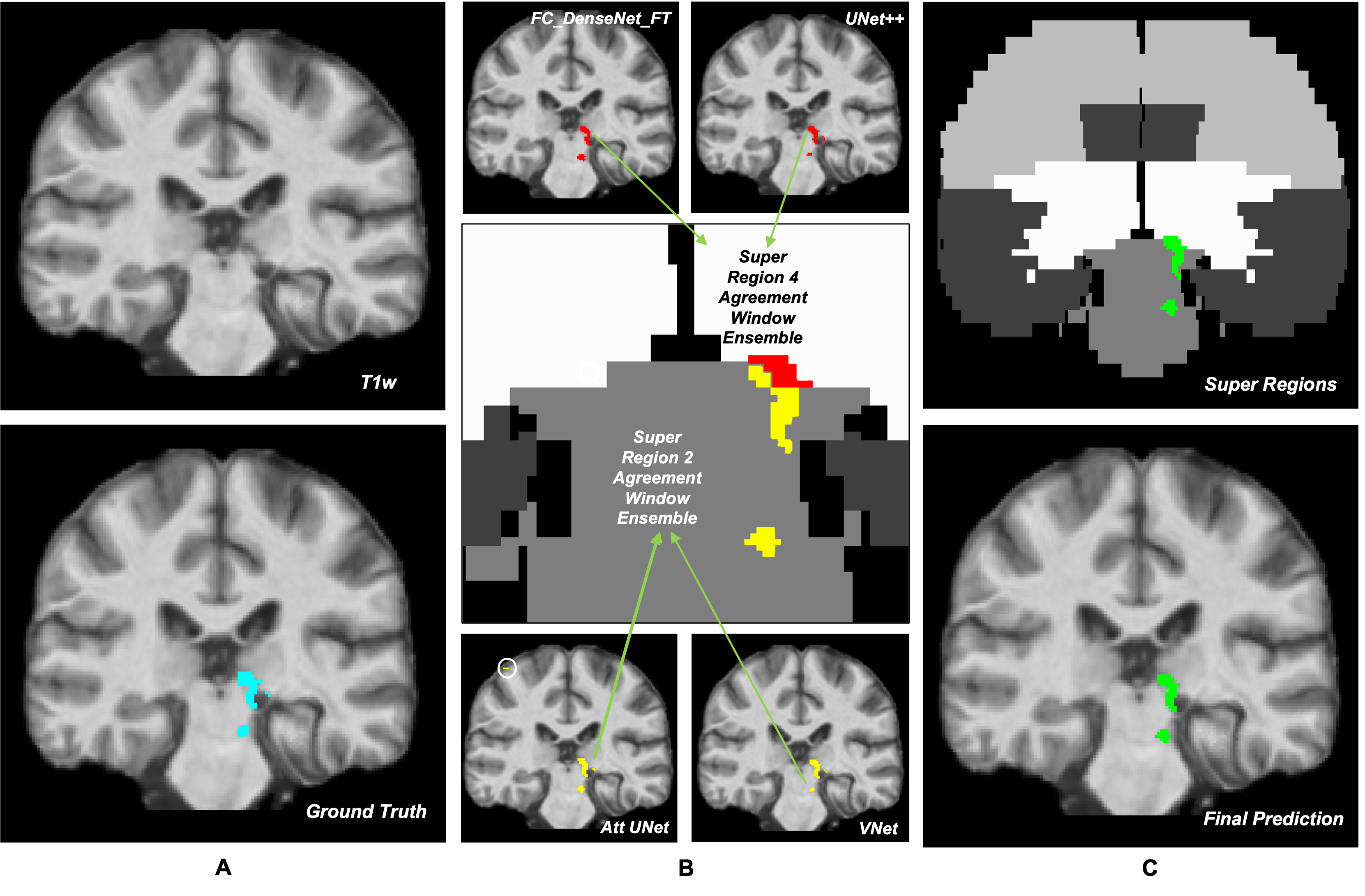}
    \caption{Comparison of hand-drawn lesion maps and predictions from distinct models and ensemble approaches. \textbf{A} represents the actual T1w MR scan and ground truth. \textbf{B} represents the predictions made by different models, and their super-region-wise ensembles. \textbf{C} represents a zoomed out view of the final super-region-wise ensemble and the subsequent final prediction superimposed on the actual T1w MR scan}
    \label{fig:4}
\end{figure}

Figure \ref{fig:4} presents a comparative analysis of the prediction results from various approaches applied to the independent validation set. For a more effective comparison, the figure displays predictions corresponding to lesions in two separate \emph{super regions}, along with their hand-drawn ground truth lesion masks. In this instance, the lesions are present in \emph{super-regions} 2 and 4. In accordance with the Table \ref{tab:2}, the models which are ensembled are Att UNet, VNet and FC DenseNet\_FT, UNet++ for \emph{super-regions} 2 and 4 respectively. The agreement window has a dimension of 2X2X2 pixels and a minimum overlap threshold of 75\% for the \emph{super-region} 4 ensemble, compared to a dimension of 3X3X3 pixels with a minimum overlap threshold of 50\% for the \emph{super-region} 2 ensemble. 

It can be observed that the agreement window ensemble methods tend to marginally overpredict the lesion area. However, it effectively predicts the targeted lesion area and exhibits almost no false positives in areas outside the lesion's location compared to other approaches.

\section{Discussion and Conclusion}

\subsection{Interpretability of results}

Our study not only offers valuable insights into the comparative performance of ensemble methods when used in conjunction with TL or mixed data training approaches but also provides a basis of comparison with other recent works using the ATLAS dataset \cite{zhou2019d, shin2022automated, verma2022automatic, bao2021mdan}. The primary goal of our research was to demonstrate that 2D-based models, when combined with TL and ensemble techniques, can achieve fast, user-experience independent, accuracy levels comparable to hand-drawn or ground truth lesion masks. 

In comparison to recent works by \cite{zhou2019d, shin2022automated, verma2022automatic}, where the ATLAS dataset is utilized for model training, the achieved Dice Similarity Coefficient (DSC) on the validation set are 0.723, 0.592, and 0.650 respectively. In our study using the ATLAS dataset, our best performing approach yields a DSC of 0.736 on the independent dataset, demonstrating the effectiveness of our method in segmenting lesions in T1w MR sequences. Furthermore, our study provides insights into lesion volume and lesion load of relevant brain structures, contributing to a more comprehensive understanding of each lesion's impact.

As we observe in Table \ref{tab:2} and \ref{tab:3}, the improvement seen using the \emph{super-region-wise} prediction can be attributed by the adaptability, reduced complexity, and noise reduction capabilities of specific model architectures, leading to more effective segmentation in particular \emph{super-regions} for some models. Furthermore, the empirically derived parameters ensure that our ensemble methods are tailored to the main aim of our current approach, namely predicting the index stroke lesion.

However, our individual models and ensemble methods have also been able to detect more wide-spread small vessel disease lesions in the stacking method as seen in Figure \ref{fig:3}A and \ref{fig:4}C highlighted circle. Upon further study, a tailored adjustment of parameters in the combined stacking and agreement window algorithm would allow us to detect small vessel disease(SVD) lesions, in particularly SVD lesions leading to T1-hypodensities \cite{wardlaw2013neuroimaging}, as well as the degeneration of long-range tracts as a secondary effects of the index lesion which can also lead to a decrease in T1 signal, but this signal decrease is seen in the anatomical course of the descending corticospinal tract (CST) \cite{devetten2010acute, yu2009longitudinal}. 

\subsection{Advantages and limitations in ensemble methods}

Three important innovations were made in our study. Along with applying TL, we implemented two ensemble methods: the stacking method and the stacking method combined with the window algorithm. A primary advantage of employing the stacking method lies in its ability to effectively reduce a significant number of false positives that frequently occur when relying solely on a single model's predictions. This improvement in accuracy can be attributed to the integration of predictions from multiple models, which consequently increases the likelihood of identifying true index lesion areas. Furthermore, the integration of the window algorithm with stacking provides an additional advantage by not only considering the overlapping predictions but also detecting edges of the lesions that might have been excluded otherwise.

However, ensemble methods also come with some limitations. The stacking method tends to underpredict lesions in some cases due to overlapping, which may lead to missed detections and a potentially incomplete assessment of the targeted lesion area. Conversely, the combined stacking and window algorithm tends to overpredict the lesions by approximately 10\% in other cases. While this overprediction can result in an inflated estimation of lesion presence, we accept this trade-off in light of the overall enhancement in model performance. Furthermore, the overprediction might be due to an increased sensitivity of the combined stacking and agreement window algorithm in detecting secondary lesions (e.g., degeneration of the corticospinal tract distal to an index lesion - which might also lead to a decrease in signal in T1w images) as well as small vessel disease lesions in areas of the brain that are unrelated to the index stroke. 

\subsection{Clinical implications and potential applications}

Our findings have considerable implications for various applications in clinical practice and in translational research, particularly in improving the accuracy and speed of diagnosis, prognosis, and treatment planning for acute, subacute and chronic stroke patients. By adopting these methods, we can improve the detection and characterization of lesions, leading to more informed decision-making,better patient outcomes predictions, as well as fast and accurate stratification of stroke patients based on lesion load data. One critical aspect of the management and treatment of stroke patients is an understanding of the impact of a lesion and correlating it to the patients' functional impairments and to their possible  gains in standard rehabilitation therapy as well as in experimental therapies. Our approach allows for the rapid and precise calculation of lesion volume and lesion load of relevant systems in the brain e.g., the CST as a surrogate structure of the motor system, enabling clinicians to evaluate the severity of the stroke, predict the likelihood of recovery as well as a posthoc determination of the failure of recovery, develop targeted rehabilitation strategies to reduce the effects of stroke-related disabilities, and provide a stratification tool for enrollment in various clinical trials. 

\subsection{Future work}

Future research work will concentrate on automating lesion segmentation and rapidly measuring the impact of a lesion on relevant brain systems utilizing machine learning approaches. Furthermore, exploring personalized treatment strategies based on individual lesion characteristics, patient demographics (e.g., biological age versus calendar age), small vessel disease lesion load, surrogate markers of brain health, and genetic factors could result in more targeted and efficient therapeutic interventions. Longitudinal studies as well as multimodal MR sequences could be pursued for fine-grain analysis of lesion progression over time.

\subsection{Conclusion}

Our study underscores the potential and relevance of applying TL to fine-tune large pre-trained 2D-based models on specific, targeted data. This approach facilitates precise segmentation of lesions in both subacute and chronic stroke patients.

The models we built for this study take into consideration unique aspects of the task at hand, focusing on lesion load and lesion volume.

A comparative analysis of different 2D model architectures was also conducted, providing valuable insights into their respective strengths and weaknesses. Our findings elucidate how variations in model architectures can influence the learning patterns of the models in different regions of the brain. This infers that different models, due to their architectural differences, can learn differently from the specific features of various brain regions.

In essence, our research not only substantiates the effectiveness of 2D-based models when fine-tuned with Transfer Learning and used in conjunction with ensemble methods, but also emphasizes the importance of choosing the appropriate model architecture for particular brain regions and task-specific criteria. Our work, therefore, offers meaningful contributions to the task of lesion segmentation in stroke patients and may pave the way for further improvements in stroke diagnosis and prognosis.

\subsection*{Supplement information}

All the tables and definitions mentioned in the manuscript are available in the supplement information. The link here is https://shorturl.at/nqIJK. 

\subsection*{Acknowledgement}

GS and AS  were partly supported by a grant from NIMH (Brain-Initiative) (7R01MH111874-05) the data analysis and computing expenses were supported by an in-kind financial contribution from Brainify, LLC. 

\subsection*{Author contributions}

Conceptualization: Sovesh Mohapatra, Advait Gosai, Gottfried Schlaug \\
Data curation: Anant Shinde, Sirisha Nouduri, Gottfried Schlaug \\
Formal analysis: Sovesh Mohapatra, Advait Gosai, Aleksei Rutkovskii \\
Funding acquisition: Anant Shinde, Gottfried Schlaug \\ 
Investigation: Sovesh Mohapatra, Advait Gosai, Aleksei Rutkovskii \\
Methodology: Sovesh Mohapatra, Advait Gosai, Gottfried Schlaug \\
Project administration: Sovesh Mohapatra, Gottfried Schlaug \\
Resources: Gottfried Schlaug \\
Supervision: Gottfried Schlaug \\
Clinical Validation: Sirisha Nouduri, Anant Shinde, Gottfried Schlaug \\ 
Writing - original draft:  Sovesh Mohapatra, Gottfried Schlaug \\
Writing - review and editing:  Sovesh Mohapatra, Advait Gosai, Gottfried Schlaug

\vfill
\pagebreak


\bibliography{sn-bibliography}

\end{document}